\documentclass[preprint]{aastex}

\slugcomment{Accepted by The Astronomical Journal for July 2000 
Publication}
\shorttitle{High Resolution Spectra of Protostars}
\shortauthors{Greene and Lada}

\begin{document}

\title{High Resolution Near-Infrared Spectra of Protostars}

% Note that \email has replaced the old \authoremail command

\author{Thomas P. Greene\altaffilmark{1}}
\affil{NASA / Ames Research Center\\ M.S. 245-6, Moffett Field, CA 
94035-1000}
\email{tgreene@mail.arc.nasa.gov}

\author{Charles J. Lada}
\affil{Smithsonian Astrophysical Observatory\\ 60 Garden Street,
Cambridge, MA 02138}
\email{clada@cfa.harvard.edu}

\altaffiltext{1}{Visiting Astronomer at the Infrared Telescope Facility
which is operated by the University of Hawaii under contract to the
National Aeronautics and Space Administration.}

\begin{abstract}
We present new high resolution ($R \simeq 21,000$) near-infrared 
($\lambda = 2 \mu$m) spectroscopic observations of a sample of Class I 
and flat-spectrum protostellar objects in the $\rho$ Ophiuchi dark 
cloud.  None of the five Class I spectra show CO v = 0 -- 2 
absorption features, consistent with high $K$-band continuum veilings, 
$4 \lesssim r_{k} \lesssim 20$ and fast stellar rotation, assuming 
that the underlying protostellar photospheres are of late spectral 
type, as is suggested by the low luminosities of most of these 
objects.  Two of the flat-spectrum protostellar objects also show no 
absorption features and are likely to be highly veiled.  The remaining 
two flat-spectrum sources show weak, broad absorptions which are 
consistent with an origin in quickly rotating ($v$ sin $i \approx 50$ 
km s$^{-1}$) late-type stellar photospheres which are also strongly 
veiled, $r_{k} \simeq 3 - 4$.  These observations provide further 
evidence that: 1)-Class I sources are highly veiled at near-infrared 
wavelengths, confirming previous findings of lower resolution 
spectroscopic studies; and 2)- flat-spectrum protostars rotate more 
rapidly than classical T Tauri stars (Class II sources) , supporting 
findings from a recent high resolution spectroscopic study of other 
flat-spectrum sources in this cloud.  In addition our observations are 
consistent with the high rotation rates derived for two of the Class I 
protostellar objects in our sample from observations of variable hard 
X-ray emission obtained with the ASCA satellite.  These observations 
suggest that certain Class I sources can rotate even more rapidly than 
flat-spectrum protostars, near breakup velocity.

\end{abstract}



\section{Introduction}

The physical natures, evolutionary states, and circumstellar disks of 
classical T Tauri stars are becoming better understood due to recent 
spectroscopic observations, high resolution imaging, and advances in 
the theory of pre-main-sequence (PMS) evolution.  However, the natures 
of the central stars and inner circumstellar environments of 
protostars are still not very well known.  This is primarily because 
they are so heavily extinguished that they are difficult to observe 
even with modern instruments and detectors.  For example, it is not 
known whether the central stars in protostellar objects differ 
substantially in effective temperature, radius, or rotation properties 
from classical T Tauri stars (CTTSs).  The presence of protostellar 
envelopes, significantly higher accretion rates, and more powerful 
outflows suggest that the photospheres of protostellar cores may 
indeed be physically different from CTTSs.  Moreover, it is not known 
whether the physical natures of protostellar photospheres are 
consistent with the predictions of protostellar evolution and PMS 
stellar theory.

Flat-spectrum and Class I young stellar objects (YSOs) are the best 
low-mass protostellar candidates for spectroscopic study because they 
have relatively well-developed central stars and are detectable at 
near-IR wavelengths.  Several pioneering studies have recently been 
undertaken to begin investigating the physical natures of these 
objects.  A few flat-spectrum protostars have been observed in low 
resolution spectroscopic surveys and thus far they appear to be 
characterized by late-type photospheres with high continuum veilings 
and sub-giant surface gravities \citep[hereafter Paper 
I]{LR99,KBTB98,CM92,PaperI}.  Even fewer have been observed at high 
spectroscopic resolution, and these observations suggest that the 
flat-spectrum protostars rotate significantly faster than the more 
evolved CTTSs \citep[hereafter Paper II]{PaperII}.

These investigations have provided some interesting clues to the 
natures of these objects, but more observations are clearly needed to 
make sense of these late-phase protostars.  It would be most 
interesting to determine if the less evolved, more heavily embedded 
and veiled Class I protostars also show near-IR photospheric 
absorption lines when observed at high resolution with high 
signal-to-noise.  Such observations could directly constrain the 
effective temperatures, gravities, veilings, and rotations of these 
objects, providing further evidence as to whether they are dominated 
by near-IR stellar, disk, or envelope emission and whether they are 
physically similar to flat-spectrum YSOs.  More flat spectrum YSOs 
should also be observed with high-resolution near-IR spectroscopy to 
confirm that they are indeed late-type rapid rotators.  In a sense the 
flat-spectrum YSOs provide a link to the well-known PMS stars, and 
this must be better developed so that they can in turn serve as a link 
to the less well-known Class I objects.

Therefore we have undertaken a new high-resolution, near-IR 
spectroscopic study of flat-spectrum and Class I YSOs in the $\rho$ 
Ophiuchi cloud core. We describe these new observations in \S 2 and 
present a rotation and veiling analysis of these data in \S 3. In \S 4 
we discuss the likely natures of each of these objects and suggest 
further observational work.

\section{Observations and Data Reduction}

Near-IR spectra were acquired in 1997 May -- June and 1998 July with 
the 3.0 m NASA Infrared Telescope Facility on Mauna Kea, Hawaii, using 
the CSHELL facility single-order cryogenic echelle spectrograph 
\citep{TTCHE90,GTTC93}.  Spectra were acquired with a 1\farcs0 (5 
pixel) wide slit on the dates indicated in Table 1, providing a 
spectroscopic resolution $R \equiv \lambda / \delta \lambda$ = 21,000 
(14 km s$^{-1}$).  The spectrograph was fitted with a 256 $\times$ 256 
pixel InSb detector array, and custom circular variable filters (CVFs) 
manufactured by Optical Coating Laboratories Incorporated were used 
for order sorting.  These filters successfully eliminated the 
significant interference fringing normally produced in CSHELL and 
other echelle spectrographs which use CVFs for order sorting.  The 
plate scale was 0\farcs20 pixel$^{-1}$ along the 30$\arcsec$ long slit 
(oriented east -- west on the sky), and all spectra were acquired at a 
central wavelength setting of 2.29353 $\mu$m corresponding to the 
v = 0 -- 2 CO band head.  Each exposure had a spectral range 
$\Delta \lambda \simeq \lambda / 400$ ($\Delta v \simeq$ 700 km 
s$^{-1}$).  Total integration times for each YSO are given in Table 1.

Data were acquired in pairs of exposures of up to 400 s duration each, 
with the telescope nodded $10\arcsec$ east or west between exposures 
so that object spectra were acquired in all exposures.  The B0V star 
HR 6165 ($\tau$ Sco) was observed periodically for telluric 
corrections.  The telescope was guided with the CSHELL internal CCD 
autoguider during exposures of these telluric correction stars, while 
the telescope tracking rates were adjusted for minimum drift while 
observing the optically invisible $\rho$ Oph YSOs.  Spectra of the 
internal CSHELL continuum lamp were taken for flat fields, and 
exposures of the internal CSHELL Ar and Kr lamps were used for 
wavelength calibrations.

All data were reduced with IRAF. First, object and sky frames were 
differenced and then divided by flat fields.  Next, bad pixels were 
fixed via interpolation, and spectra were extracted with the APALL 
task.  Extracted spectra were typically 5 pixels (1\arcsec) wide along 
the slit (spatial) direction at their half-intensity points.  Spectra 
were wavelength calibrated using low-order fits to lines in the arc 
lamp exposures, and spectra at each slit position of each object were 
co-added.  Instrumental and atmospheric features were removed by 
dividing wavelength-calibrated object spectra by spectra of early-type 
stars observed at similar airmass at each slit position.  Final 
spectra were produced by combining the spectra of both slit positions 
for each object.

\section{Data Analysis and Results}

\subsection{Object Sample}

The object sample was selected from the Class I and flat-spectrum YSOs 
observed at low spectral resolution ($R \simeq 500$) in Paper I which 
were not subsequently observed at high spectral resolution in Paper 
II. None of the newly observed sources (listed in Table 1) showed any 
absorption features in their low resolution $K$-band spectra (Paper 
I), and they are also relatively bright, 7 mag $\lesssim K \lesssim$ 
10 mag.  Table 1 shows that the brighter point sources (i.e. Elias 29, IRS 
54) were observed with higher signal-to-noise ratios than the fainter 
ones (i.e. IRS 43, WL 6).  This is useful for analyzing the veiling in 
these objects if the brightness differences among the sources are 
mostly due to different amount of IR excess emission from 
circumstellar regions.  If this is true, then the brighter objects 
have greater near-IR veiling and greater signal-to-noise is required 
to detect their photospheric absorption lines.  On the other hand, it 
may be possible that some of the bright protostars are featureless 
because they are of relatively early spectral type (G or earlier) and 
possess intrinsically weak $K$-band absorption lines (besides H and 
He).

\subsection{Veiling and Rotation Analysis}

The new flat-spectrum and Class I YSO spectra are shown in Figures 1 
and 2, respectively.  None of the five Class I spectra show any 
evidence of CO absorption.  Two of the flat-spectrum YSOs also show no 
evidence of any CO absorption (GSS 26 and YLW 13B), while the other 
two show evidence of weak, broad band heads and perhaps some 
overlapping rotation-vibration lines as well (Figure 1).

We now analyze these spectra to constrain the veilings and physical 
natures of these sources.  Our previous high resolution study (Paper 
II) showed that late-type Class II YSOs (pre-main-sequence stars) 
rotate slowly, $\langle v {\rm sin} i \rangle < 20$ km s$^{-1}$, while 
flat-spectrum ones rotate quickly, $\langle v {\rm sin} i \rangle > 
20$ km s$^{-1}$.  This study also showed that in addition to 
broadening the band head, high rotation also decreased the maximum 
absorption depth of the band head and adjacent individual 
rotation-vibration lines.  Thus rotation as well as veiling can reduce 
the detectability of CO absorption in finite signal-to-noise spectra.

We now explore the limits of $K$-band veiling and rotation in our new 
protostar sample by comparing their spectra to those of flat-spectrum 
and Class II YSOs to which we have artificially added continuum 
veiling.  We chose VSSG 25 to be representative of a slowly rotating 
Class II YSO ($v$ sin $i = 5$ km s$^{-1}$) and VSSG 17 to be 
representative of a quickly rotating ($v$ sin $i = 47$ km s$^{-1}$) 
flat-spectrum source (see Paper II).  The spectral types and $K$-band 
veilings have been measured for both of these objects.  VSSG 25 has a 
spectral type M0IV/V with $r_{k}$ = 0.25 \citep{LR99} while VSSG 17 is 
M0IV/V with $r_{k}$ = 0.9 (\citeauthor{LR99}; Paper I).  The $K$-band 
veiling is defined as $r_{k} = F_{Kex}/F_{K*}$ where $F_{Kex}$ 
is the $K$ band excess flux, and $F_{K*}$ is the $K$ band stellar flux.
We added veiling (a constant positive offset) to the template spectrum 
of VSSG 17 (taken from Paper II) and degraded its signal-to-noise so 
that its CO absorption equivalent width and detectability were 
weakened.  We then used a series of these VSSG 17 templates with 
various veilings and signal-to-noise ratios to estimate the veilings 
of our observed sources.

We estimated the veiling for sources which show CO band head 
absorptions (IRS 51 and IRS 63) by matching their spectra to veiled 
templates with identical signal-to-noise and similar features (CO band 
head depth and slope).  The band head profiles of these objects also 
matched the shape of VSSG 17 and matched those of observed slowly 
rotating late-type stars which had been artificially broadened with a 
stellar rotation profile of $v$ sin $i = 50$ km s$^{-1}$ (see Paper 
II).  Thus it is likely that their CO absorptions arise in rapidly 
rotating stellar photospheres.  However, the absorption features of 
these highly veiled sources are very weak, and this limits the 
derived rotation velocities to uncertainties of approximately 40\% and 
the derived veilings to uncertainties of at least 20\%.  

The minimum likely veilings of the Class I and flat-spectrum sources 
without detectable CO absorptions were estimated by assuming that 
their CO absorptions were intrinsically similar to VSSG 17 but had 
increased continuum veiling.  We derived an analytical relation for 
veiling based on the principle that an object's CO band head is 
undetectable when its maximum absorption depth is less than 3.0 times 
the RMS noise over 1 resolution element (5 pixels) in the spectrum.  
This is a robust minimum likely veiling criterion because less 
veiling would ensure that the band head would be definitely detected, 
and considering more resolution elements would increase the amount of 
veiling derived.  The resultant derived minimum likely $r_{k}$ values 
matched those of templates which were veiled to the point where their 
CO absorptions just disappeared visually.

Possible differences between the actual spectral types of these 
sources and the M0 template also cause uncertainties in the derived 
veilings of up to about 50\%, but most deviations are expected to be 
smaller than this.  All observed Class I sources except Elias 29 are 
relatively low luminosity, 2.4 L$_{\odot} \leq$ L$_{\rm bol} \leq 13$ 
L$_{\odot}$ \citep[hereafter WLY]{WLY89}.  All observed flat-spectrum 
YSOs have 1 L$_{\odot} \leq$ L$_{\rm bol} < 3$ L$_{\odot}$ (WLY; 
\citeauthor{GWAYL94} \citeyear{GWAYL94}).  These luminosities are 
consistent with the sources being low-mass PMS YSOs (M $<$ 1 
M$_{\odot}$) if the Class I objects are powered by accretion, an 
assumption which is consistent with their high veilings (see also \S 
4.2).  In using VSSG 17 as a spectral type template to measure 
veilings, we have implicitly assumed that all sources have spectral 
types near M0IV/V, a typical value for T Tauri stars.  A young star of 
this spectral type has a mass of approximately 0.4 M$_{\odot}$ if on 
the birthline of the H--R Diagram \citep[see][]{S88}.  If an observed 
YSO is really 1 M$_{\odot}$, then it would have a spectral type of 
K3--4IV/V \citep[see][]{DM97, S88} and thus an intrinsic stellar CO 
equivalent width of only about 0.6 times that of an M0IV/V star (see 
Paper I).  This would mean that its actual veiling is $r'_{k} = 0.6 
r_{k} - 0.4$ (see \S 4.2 of Paper I) where $r_{k}$ is the veiling 
estimate made assuming an M0IV/V spectral type.  Likewise, if an 
observed object were really a lower mass star near the brown dwarf 
limit, then its true birthline spectral type would be near M6IV/V and 
its intrinsic CO absorption would be about 40\% greater than that of 
the M0IV/V template.  Thus the true veilings of observed sources may 
differ from our derived ones by as much as 50\%.  However, it is 
unlikely that many of the observed objects have such large deviations 
(in either direction) because their luminosities are consistent with 
most having masses of approximately 0.5 M$_{\odot}$, similar to that 
expected for our M0IV/V template.  We have used other published 
information on the observed objects when available to estimate their 
intrinsic spectral types before deriving veiling estimates (see \S 4.1 
and 4.2).

Finally, we studied how rotation decreases the apparent CO band head 
absorption depths.  This effect can be separated from veiling by 
measuring the slope of the CO band head, but this is not possible for 
sources which do not show this absorption feature.  The majority of 
sources in our sample do not show this feature, so we do not know 
their rotation velocities.  Our earlier study (Paper II) showed that 
flat-spectrum sources rotate significantly faster than Class II YSOs, 
so our current sample of more highly embedded flat-spectrum and Class 
I YSOs may rotate even faster still (perhaps $v$ sin $i > 50$ km 
s$^{-1}$).  Indeed, recent X-ray observations with the ASCA satellite 
indicate that one Class I source in our sample (IRS 43) is rotating at 
least this rapidly \citep{MGTK00}.  We estimate that the 
equatorial rotational breakup velocities of these young stars ($M 
\simeq 0.5 M_{\sun}$ and $R \simeq 3 R_{\sun}$) are approximately $v 
\simeq 180$ km s$^{-1}$, or $v$ sin $i \simeq 150$ km s$^{-1}$ for a 
mean inclination $i = 57\deg$.  Next, we studied how rotation 
decreases the maximum CO band head absorption in YSOs by artificially 
rotating our WL 5 template by convolving its observed spectrum with 
limb-darkened stellar broadening profiles for 25 km s$^{-1} \leq v$ 
sin $i \leq 175$ km s$^{-1}$.  See Paper II for more details and 
examples of artificially rotated spectra.

These experiments showed that the maximum CO absorption depth of the 
VSSG 25 template with $v$ sin $i = 150$ km s$^{-1}$ is a factor of 
1.37 weaker than the one with $v$ sin $i = 50$ km s$^{-1}$.  Therefore 
if VSSG 17 were rotating at breakup, its maximum CO absorption depth 
would be only 73\% as deep as now seen in its spectrum.  Thus if 
rotating near breakup, the featureless objects in our sample would 
have somewhat lower continuum veilings than those calculated based on 
the observed VSSG 17 spectrum.  We have calculated these reduced 
values, and in Table 2 we present all of these estimated veilings for 
the observed sources, using our best estimates of their intrinsic 
spectral types.

\section{Discussion and Conclusions}

We now discuss how the results of this veiling / rotation analysis and
pre-existing data constrain the possible physical natures of these 
sources. 

\subsection{Flat-Spectrum Objects}

The flat-spectrum YSOs IRS 63 and IRS 51 were both found to have 
broad, weak CO absorptions which matched those expected for late-type 
stellar photospheres rotating at $v$ sin $i \simeq 50$ km s$^{-1}$.  
The weak CO absorptions of these two YSOs are consistent with their 
not being detected in our initial low-resolution survey (Paper I).  We 
estimate the continuum veiling of IRS 63 to be $r_{k} \simeq 4$ 
provided that it is a PMS YSO near M0 spectral type.  
\citeauthor{LR99} find that the spectral type of IRS 51 is G5 -- K7, 
earlier than our M0 template VSSG 17.  Thus their derived veiling 
$r_{k}$ = 1 -- 3 is lower than ours because a G5 -- K7 PMS star has 
less intrinsic CO absorption than a M0 one (see \S 3.2).  It is likely 
that IRS 51 is indeed an embedded low-mass YSO because its bolometric 
luminosity is only 1.4 L$_{\odot}$ (WLY).  The birthline mass for this 
luminosity is approximately 0.5 M$_{\odot}$, corresponding to a 
spectral type of K5--7 and a true veiling of $r_{k} \simeq 3$.  Thus 
it is likely that both IRS 63 and IRS 51 are similar to the quickly 
rotating flat-spectrum YSOs which we analyzed in Paper II, but these 
new objects have even greater veiling (i.e. $r_{k}$ = 3 -- 4 versus 
$r_{k} \simeq 1$ for the Paper II YSOs).

\citeauthor{LR99} found GSS 26 to have variable veiling, $r_{k}$ = 
0.75 and $r_{k}$ = 4 at epochs of 1994 July and 1996 May, 
respectively.  Our spectrum of GSS 26 (in Figure 1) was taken in 1997 
May, one year after the latest \citeauthor{LR99} spectrum.  We 
estimate that $r_{k} \gtrsim 11$ when our spectrum was acquired, and 
our assumption of a M0 spectral type is consistent with the 
\citeauthor{LR99} determination of K5 -- M2.  This rapid increase in 
veiling - a factor of 2 each year - is perhaps suggestive of a 
similarly rapid increase in accretion.  \citeauthor{LR99} also note 
that this source increased in brightness by $K \simeq$ 1.2 mag between 
epochs.  Our spectra are not photometrically calibrated, but 
comparisons with other objects support that this source was at least 
as bright as when observed by \citeauthor{LR99} the previous year.  
This is one of the YSOs \citeauthor{LR99} observed whose HI Br 
$\gamma$ emission line flux increased as its veiling increased, 
implying that the excess continuum emission is associated with a 
circumstellar accretion disk if the Br $\gamma$ emission arises from 
disk accretion.  The rapid variability of this object's veiling also 
suggests that its $K$-band veiling is produced by accretion from inner 
disk distances (several AU) and not from an outer disk or an outer 
circumstellar envelope.

The high $r_{k}$ values of these objects also constrain the physical 
origins of their veilings.  In Paper I we showed that veilings $r_{k} 
> 1$ cannot be produced by a simple optically thick, geometrically 
thin reprocessing disk around a low-mass PMS star.  Consequently we 
argued that these high veilings are most likely produced by either 
actively accreting circumstellar disks or circumstellar envelopes 
associated with these objects.  Furthermore we found that veilings in 
the range measured for IRS 51 and IRS 63, $r_{k}$ = 3 -- 4, can be 
caused by luminous accretion disks ($L_{disk}/L_{*} \leq 3$).  
Veilings in the range observed for GSS 26, $r_{k} \simeq 5 - 10$, 
could be explained by extremely luminous accretion disks 
($L_{disk}/L_{*} \geq 3$).  In either case, these accretion disks 
would have to have relatively large central holes to avoid producing 
strong CO absorption-line systems in the disk photosphere itself.  
However, because there is no obvious physical mechanism for producing 
central holes of the needed size, \citet{CHS97} suggested that the 
veiling flux must originate in some other circumstellar structure such 
as the inner regions of the protostellar envelope.  On the other hand, 
this seems to be inconsistent with the observation by 
\citeauthor{LR99} that the $K$-band excesses of flat-spectrum YSOs are 
correlated with their HI Br $\gamma$ line fluxes which in turn 
suggests that the veiling flux should originate in the disk.  More 
detailed knowledge of the conditions required to produce CO absorption 
line systems in an accretion disk may be needed to resolve this issue.

The flat-spectrum source YLW 13B was found to have H Br $\gamma$ 
absorption by \citeauthor{LR99}, who estimate its spectral type to be 
earlier than K0.  However, they also find it to be significantly 
veiled with $r_{k} > 1$, so it is possibly an intermediate mass PMS 
cloud population member.  Our non-detection of CO absorption does not 
constrain this source further.

\subsection{Class I Objects}

We do not detect CO absorptions in any of the Class I YSOs which we 
observed (Figure 2), confirming earlier low resolution spectroscopic 
observations that found all these objects to be featureless and highly 
veiled (Paper I; \citeauthor{LR99}).  Consequently their spectral 
types are unknown, however all of these Class I sources exhibit HI Br 
$\gamma$ emission (Paper I; \citeauthor{LR99}).

Analysis of our new data constrains the natures of these objects.  
Table 2 shows that they all have large veilings, $r_{k} > 4$, if they 
are late-type ($\sim$ M0) stars.  GSS 30, IRS 43, and WL 6, all have 
estimated minimum veilings of $r_{k} \simeq 5-8$, overlapping with the 
flat-spectrum sample.  These sources all have bolometric luminosities 
L$_{\rm bol} \leq 13$ L$_{\odot}$ (WLY).  This is consistent with 
their being low-mass (M $<$ 1 M$_{\odot}$) protostars accreting matter 
at rates $\dot{M} \simeq 5 \times 10^{-6}$ M$_{\sun}$ yr$^{-1}$, the 
value expected for the T $\simeq$ 20 K gas temperatures in the $\rho$ 
Oph cloud \citep[see][hereafter ALS]{ALS87}.  Veilings in the observed 
range, $r_{k} > 4 - 10$, are also predicted by theoretical models of 
Class I circumstellar envelopes (Paper I; Calvet et al.).

\citeauthor{MGTK00} note that the maximum possible mass of IRS 43 
(also known as YLW 15) is 2.2 M$_{\odot}$ which is derived from PMS 
models given its bolometric luminosity (L $\sim$ 10 L$_{\odot}$) and 
assuming it is on the birthline.  Likewise, they calculate that the 
maximum likely mass of WL6 is approximately 0.4 M$_{\odot}$; this 
increases somewhat if the L$_{\rm bol}$ = 2.4 L$_{\odot}$ of WLY is 
adopted.  These maximum calculated masses assume that essentially all 
luminosity is due to photospheric thermal radiation ($L = 4\pi 
R^{2}\sigma T^{4}$) and essentially none is due to accretion ($L = G 
M\dot{M}/R$).  These sources are discussed further in $\S 4.3$.

We estimate that Elias 29 and IRS 54 have very high veilings if they 
are late-type low-mass stars, $r_{k} > 14 - 34$.  The bolometric 
luminosity of IRS 54 is estimated to be only 12 L$_{\odot}$ (WLY), 
also consistent with this object being a low-mass protostar which is 
accreting its envelope at the rate prescribed by the $\rho$ Oph 
cloud's gas temperature.  However, its continuum veiling must be 
$r_{k} > 14 - 20$ if it has spectral type M0 and is rotating rapidly.  
This is about a factor of 2 higher than the model predictions of 
Calvet et al., but those calculations were done for a hypothetical 
$\rho$ Oph Class I YSO with L = 5 L$_{\odot}$.  The model may predict 
greater veiling for IRS 54 if its higher luminosity is taken into 
consideration.  It is also possible that this source may be a somewhat 
earlier type protostar which is less veiled.

Elias 29 has the highest derived veiling of the sample, $r_{k} > 25 - 
34$, assuming an intrinsic M0 photosphere and a high rotation rate.  
WLY estimate its luminosity to be L$_{\rm bol}$ = 48 L$_{\odot}$, and 
ALS have modeled it as a 1 M$_{\odot}$ protostar which is accreting 
its circumstellar envelope.  Such a star would have a spectral type of 
K3--4 if on the birthline \citep[see][]{DM97, S88}, with an intrinsic 
CO absorption approximately 60\% as strong as that of an M0 star (see 
\S 3.2).  Thus we revise our estimate of the likely veiling of this 
YSO to $r_{k} > 15 - 20$ if it is indeed a 1 M$_{\odot}$ protostar.  
In Paper I we analyzed the ALS model for Elias 29 and showed that the 
predicted emission from the inner protostellar envelope of this source 
would produce a veiling of $r_{k} \approx 20$, assuming that its disk 
luminosity is 0.75 L$_{\rm bol}$.  Thus our new measurement is 
consistent with our earlier prediction based on the ALS model.

The high luminosity of Elias 29 also allows for it being a more 
massive, earlier spectral type YSO that has higher stellar luminosity 
and less accretion luminosity than assumed by the ALS model.  However, 
it is unlikely to be very different because the observed near-to-far 
IR energy distribution and the 10 $\micron$ silicate absorption of 
Elias 29 are fit well by the ALS model, and there are no clues which 
indicate that Elias 29 is an early-type object.  For example, WL 16, 
which is likely an early A type star \citep{BHCRL00}, has mid-IR 
aromatic hydrocarbon emission features which indicate a UV radiation 
field \citep{HTG92}.  However, Elias 29 shows no evidence for IR 
hydrocarbon emission \citep{HBT95,B99} and thus no evidence for a UV 
radiation field.

\subsection{Protostellar Rotation}

Our observations, specifically the broad band head shapes of IRS 51 and 
IRS 63, strengthen the earlier findings of Paper II which suggested 
that flat-spectrum protostars rotate more rapidly than Class II 
sources (CTTSs).  Since flat-spectrum protostars are believed to be 
evolutionary precursors of CTTSs, then this finding may indicate that 
certain physical conditions characteristic of protostellar evolution 
(e.g., high accretion rates) may result in their higher rotation 
rates.  It is therefore interesting to ask whether the less evolved 
Class I sources might rotate even more rapidly than the flat spectrum 
sources.  Indeed, some models of protostar development predict that 
such objects should be rotating near breakup \citep{S91}.

Recently, strong hard X-ray flares have been observed with the ASCA 
satellite from two of the Class I protostars in our sample -- IRS 43 
and WL 6 (\citeauthor{MGTK00}).  These remarkable observations reveal 
relatively rapid periodicities in the X-ray emission from these 
sources, enabling the derivation of their photometric rotation rates.  
\citeauthor{MGTK00} find that WL 6 is rotating with a period of about 3 
days ($v$ sin $i \simeq 40$ km s$^{-1}$ for a 0.5 M$_{\sun}$ star on 
the birthline), comparable to the rotation rates of the flat-spectrum 
sources observed here and in Paper II. This source also has a weak 
outflow \citep{STUKTH97}, undetected millimeter emission from its 
envelope \citep{AM94}, and an IR energy distribution which can be 
modeled as a highly extinguished flat spectrum YSO 
(\citeauthor{MGTK00}).  All of these properties indicate that WL 6 may 
indeed be very similar to the flat spectrum YSOs for which we have 
detected absorption lines and have found to be rotating more rapidly 
than CTTSs (Class II YSOs).

\citeauthor{MGTK00} also argue that the central star of IRS 43 has a 
20 h rotation period, the observed period of its X-ray variability.  
This requires that its mass be greater than or equal to 1.8 
M$_{\odot}$ in order for it to be rotating below breakup velocity if 
it is on the birthline (\citeauthor{MGTK00}).  Thus the mass of IRS 43 
is constrained to lie in the range 1.8 -- 2.2 M$_{\odot}$ by its X-ray 
emission and bolometric luminosity (see $\S 4.2$) if its rotation 
period is indeed equal to its X-ray variability period of 20 h.  IRS 
43 does have a more steep mid-IR energy distribution (clearly Class I) 
than WL 6, and it also has spatially-resolved (r $\simeq$ 3000 AU) 
millimeter emission with a derived envelope mass of approximately 0.1 
M$_{\odot}$ (\citeauthor{AM94}).  Therefore it is likely to be in an 
earlier evolutionary state than WL 6.

The slow rotation velocities of CTTSs have been explained by angular 
momentum regulation of these stars by magnetic coupling to their 
disks.  \citet{ESHSHHAMPG93} found that late-type T Tauri stars (TTSs) 
with large $H-K$ IR excesses (CTTSs) had slow rotation periods, P $>$ 
4 d.  They also found that TTS with small $H-K$ IR excesses had a 
broad range of periods, including a significant number with P $<$ 4 
days.  \citeauthor{ESHSHHAMPG93} interpreted this correlation to arise 
because the magnetic fields of the CTTSs were were coupled to their 
disks, providing stellar angular momentum regulation and therefore 
long stellar rotation periods.  The low-excess TTSs had already 
dissipated their disks and so were not subject to this regulation 
mechanism.  More recent studies of larger TTS samples have both 
disputed that the correlation between IR excess and rotation period 
exists \citep{SMMV99} and have provided evidence that it exists but is 
weak \citep{HRHC00}. 

Our studies of YSO rotation (this paper and Paper II) have shown that 
flat-spectrum protostars rotate significantly more rapidly than Class 
II YSOs or CTTSs, suggesting that rotation velocities decrease as 
stars evolve past the protostellar state.  This scenario has been 
bolstered and expanded further by the recent X-ray results of 
\citeauthor{MGTK00}.  Taken together (and along with the many rotation 
studies of optically visible CTTSs), these works suggest that heavily 
embedded protostars (Class I) rotate very rapidly, in some cases near 
breakup velocity, while less embedded ones (flat--spectrum YSOs) 
rotate somewhat less rapidly, at about 1/3 breakup velocity ($v$ sin 
$i \simeq 50$ km s$^{-1}$), and Class II YSOs / CTTSs rotate slowly, 
$v$ sin $i < 20$ km s$^{-1}$. This finding would be strengthened 
considerably by further cross-checking of observational techniques; 
the X-ray protostars should be observed at higher signal-to-noise in 
the near-IR to search for rotationally broadened lines, while the flat 
spectrum and Class II YSOs with IR-derived rotation velocities should 
be observed for periodic X-ray variability.

If the angular momenta of low-mass YSOs are indeed regulated by 
star--disk coupling, then the fact that flat-spectrum (and at least 
one Class I) YSOs rotate significantly more rapidly than CTTS implies 
that either the flat-spectrum / protostellar YSOs are coupled to 
faster rotating disk regions than CTTSs, or else that stars and disks 
do not become rotationally locked until the CTTS evolutionary phase.  
In the first case, flat-spectrum and Class I YSOs may couple to their 
disks at smaller radii (and hence have higher rotation velocities) 
because their accretion rates are much higher than CTTS. The veilings 
and luminosities of Class I and flat-spectrum YSOs are considerably 
higher (by about an order of magnitude) than CTTS, supporting the 
notion that they have higher accretion rates also.  In support of the 
second case, \citeauthor{MGTK00} have posited that protostars are not 
initially magnetically coupled to their disks but rather spin-down and 
become coupled over a magnetic braking time on the order of 10$^{5}$ 
yr which is nearly linearly proportional to stellar mass.  This is 
comparable to the lifetime of the Class I and flat spectrum phases, so 
this would account for the higher rotation velocities of flat-spectrum 
and Class I protostars.  This latter magnetic braking scenario of 
velocity evolution from Class I to flat-spectrum to CTTS YSOs may be 
somewhat complicated by mass effects; \citeauthor{MGTK00} predict that 
at the same age more massive protostars will rotate more quickly than 
less massive ones.  Protostellar masses must be measured much more 
accurately before this effect can be verified, however.
  
Obtaining new high resolution, high signal-to-noise spectra over the 
entire 1.5 -- 2.4 $\micron$ region will likely be the best method for 
obtaining more definitive information on the masses (spectral types) 
and rotational characteristics of Class I protostars.  This wide 
spectral range is required in order to be sensitive to a wide range of 
spectral types.  Even intermediate-to-high mass stars with HI line 
emission may show $H$-band HI Br absorption lines which may strongly 
constrain spectral types and masses (e.g., \citeauthor{BHCRL00}), 
while $K$-band data are required to determine the properties of very 
red late-type YSOs.  Detecting and resolving near-IR lines in WL 6 and 
IRS 43 would allow determination of their masses and photospheric 
rotation rates, providing a good test of the emerging scenario of 
protostellar rotational evolution.

\acknowledgments

We thank the referee John Lacy and also Pat Cassen for providing 
comments which improved this paper.  We also thank W. Golisch, D. 
Griep, and C. Kaminski for assistance with the observations.  We 
acknowledge the National Science Foundation for funding grant 
AST--9420506 to develop the fringe-free CVFs used to acquire these 
data with CSHELL. TPG acknowledges a grant from the NASA Ames Research 
Center Director's Discretionary Fund.  All data were reduced with 
IRAF, which is distributed by the National Optical Astronomy 
Observatories, which is operated by the Association of Universities 
for Research in Astronomy, Inc., under contract to the National 
Science Foundation.



%% Tables should be submitted one per page, so put a \clearpage before

%% Two options are available to the author for producing tables:  the
%% deluxetable environment provided by the AASTeX package or the LaTeX
%% table environment.  Use of deluxetable is preferred.
%%

\clearpage

%% Tables may also be prepared as separate files. See the accompanying
%% sample file table.tex for an example of an external table file.
%% To include an external file in your main document, use the \input

%% No more than seven \figcaption commands are allowed per page,
%% so if you have more than seven captions, insert a \clearpage

%% There must be a \figcaption command for each legend. Key the text of the
%% legend and the optional \label in curly braces. If you wish, you may
%% include the name of the corresponding figure file in square brackets.
%% The label is for identification purposes only. It will not insert the
%% figures themselves into the document.
%% If you want to include your art in the paper, use \plotone.
%% Refer to the on-line documentation for details.

\begin{deluxetable}{lrrlrrl}

\tablenum{1}
%\tablewidth{33pc}
\tablewidth{0pc}
\tablecaption{IRTF CSHELL 2.2935 $\micron$ CO Observations}
\tablehead{
\colhead{Source}      & 
\colhead{$\alpha$(1950)} & \colhead{$\delta$(1950)} &
\colhead{UT Date} &
\colhead{Int. Time}   & \colhead{S/N\tablenotemark{a}} &
\colhead{Features} \\ [0.2 ex]
\colhead{}            & 
\colhead{\small hh mm ss.s} & \colhead{$\arcdeg$  $\arcmin$  $\arcsec$} &
\colhead{} &
\colhead{(minutes)}   & \colhead{} &
\colhead{}
}
\tablecolumns{7}
\startdata

\sidehead{$\rho$ Oph Class I YSOs:}

GSS 30	  & 16 23 20.0 & -24 16 22 & 1999 Jul 07 & 74.0 &  25 &  none\\
Elias 29  & 16 24 07.8 & -24 30 35 & 1997 Jul 07 & 46.0 & 135 &  none\\
WL 6      & 16 24 20.0 & -24 23 11 & 1998 Jul 04 & 102  &  25 &  none\\
IRS 43    & 16 24 25.1 & -24 34 10 & 1997 May 05 & 56.7 &  35 &  none\\
IRS 54    & 16 24 50.0 & -24 25 05 & 1997 May 05 & 63.3 &  80 &  none\\
\\

\sidehead{$\rho$ Oph Flat-Spectrum YSOs:}

\\
GSS 26    & 16 23 09.0 & -24 14 11 & 1997 May  5 & 30.0 &  65 & none\\
YLW 13B   & 16 24 19.3 & -24 35 03 & 1997 May  5 & 40.0 &  65 & none\\
IRS 51    & 16 24 37.6 & -24 36 35 & 1997 Jul  4 & 60.0 &  50 & weak broad abs\\
IRS 63    & 16 28 34.3 & -23 55 05 & 1997 Jul  4 & 43.3 &  40 & weak broad abs\\

\tablenotetext{a}{Approximate signal-to-noise ratios were determined in 
relatively featureless regions of spectra in regions of good atmospheric
transmission.}

\enddata
\end{deluxetable}

\clearpage

\begin{deluxetable}{lrrl}
\tablenum{2}
\tablewidth{28pc}
\tablecaption{Derived Veiling Estimates}
\tablehead{
\colhead{Source}    & \colhead{$r_{k}$\tablenotemark{a}} &
\colhead{$r_{k,\rm break}$\tablenotemark{b}}   & \colhead{Comment}
}
\tablecolumns{4}
\startdata
\sidehead{$\rho$ Oph Class I YSOs:}

GSS 30	  &   5.5 & 3.8 & lower limits\\
Elias 29  &   20\tablenotemark{c} &  15\tablenotemark{c} & lower limits\\
WL 6      &   5.5 & 3.8 & lower limits\\
IRS 43    &   8.1 & 5.7 & lower limits\\
IRS 54    &    20 & 14  & lower limits\\

\sidehead{$\rho$ Oph Flat-Spectrum YSOs:}
GSS 26    &  16 & 11 & lower limits\\
YLW 13B   &  \nodata & \nodata & \tablenotemark{d}\\
IRS 51    &   3\tablenotemark{e} &    &  measured value\\
IRS 63    &   4 &    & measured value\\

\tablenotetext{a}{\small Minimum likely $K$-band veiling assuming 
intrinsic CO depths and rotation velocities similar to VSSG 17. See 
\S 3.2 for uncertainties.}

\tablenotetext{b}{\small As for note (a) but computed for sources 
rotating near breakup velocity.}

\tablenotetext{c}{The minimum veilings shown were computed for a 1 
M$_{\sun}$ object, with intrinsic CO absorptions 40\% weaker than VSSG 
17. (see \S 4.2}

\tablenotetext{d}{\small We do not report our derived veilings since 
this is clearly not a late-type object (Luhman \& Rieke 1999; see 
\S4.1).}

\tablenotetext{e}{\small The Luhman \& Rieke (1999) spectrum and 
L$_{\rm bol}$ = 1.4 L$_{\odot}$ of IRS 51 constrain its spectral type 
to K5 -- K7, which we have assumed to derive its veiling (see \S 4.1).}

\enddata

\end{deluxetable}

\begin{figure}
    \figurenum{1} 
    \plotone{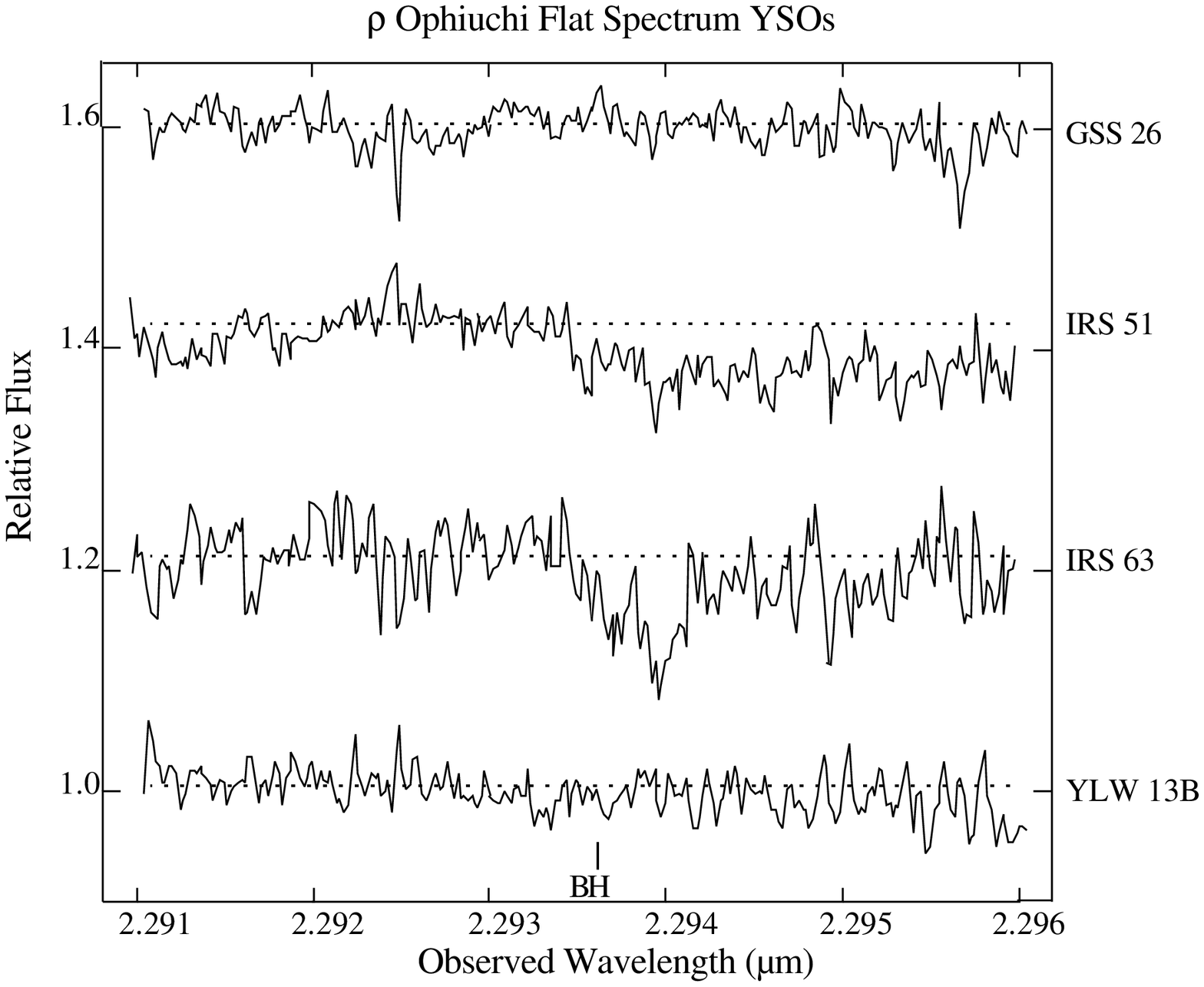}     
    \caption{Spectra of flat-spectrum $\rho$ Oph protostars in the v = 
    0 -- 2 CO band head region.  The approximate location of the 
    2.2935 $\mu$m band head is indicated (BH), and the approximate 
    continuum level of the region red-ward of the band head is shown 
    by a dotted line for each spectrum.  Each spectrum has been scaled 
    to a mean value of 1.  YLW 13B has a zero point of 0, but the 
    other spectra are offset by a value of 0.2 relative to each other.  
    The sources IRS 51 and IRS 63 show clear evidence for CO band head 
    absorptions, while GSS 26 and YLW 13B do not.  The spectral 
    resolution element is approximately $1.1 \times 10^{-4} \micron$, 
    so the band heads of IRS 51 and IRS 63 are clearly broadened by 
    rotation.}
\end{figure}

\begin{figure}
    \figurenum{2} 
    \plotone{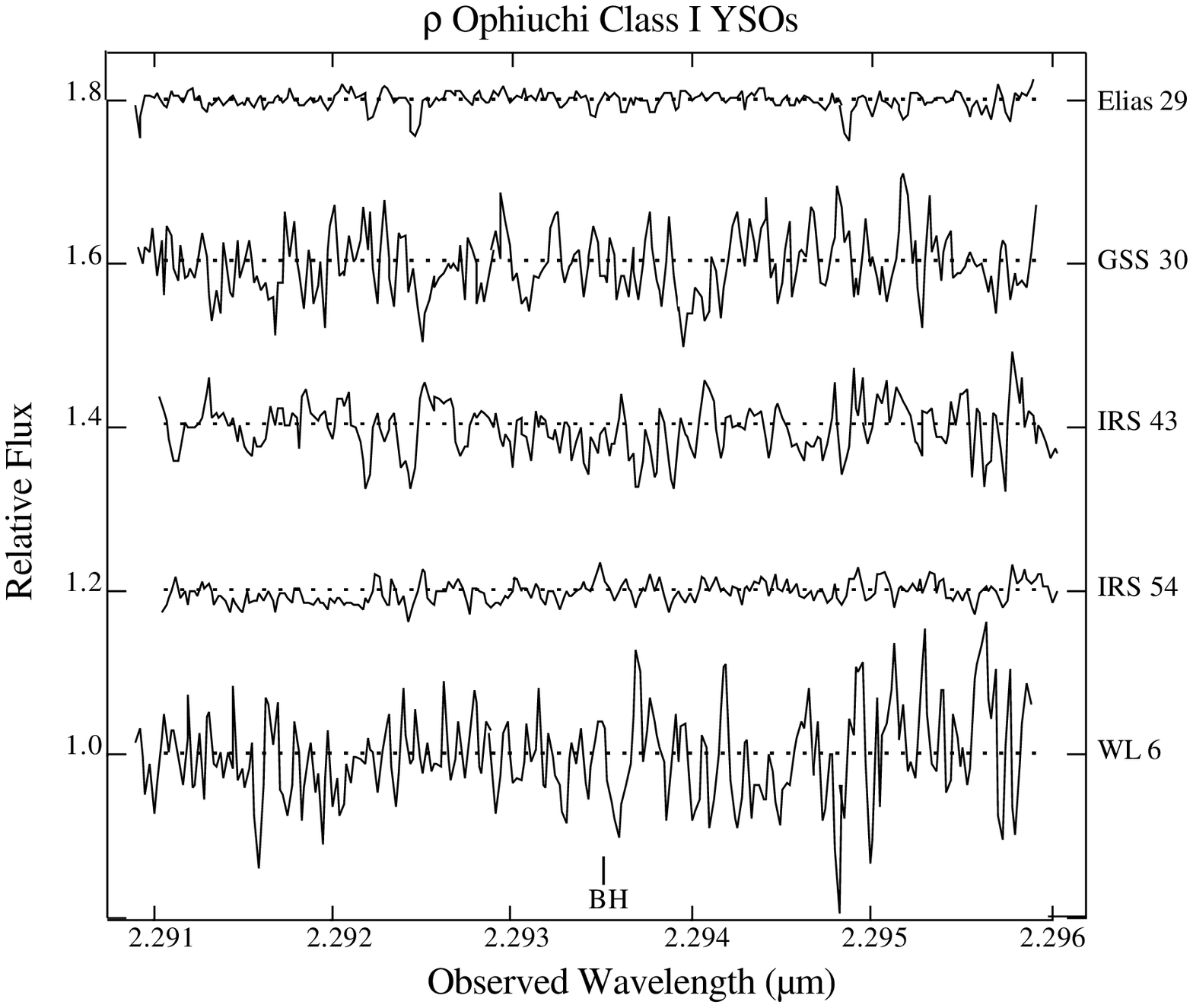} 
    \caption{Spectra of Class I $\rho$ Oph protostars in the v = 0 -- 
    2 CO band head region.  The approximate location of the 2.2935 
    $\mu$m band head is indicated (BH), and the approximate continuum 
    level of the region red-ward of the band head is shown by a dotted 
    line for each spectrum.  Each spectrum has been scaled to a mean 
    value of 1.  WL 6 has a zero point of 0, but the other spectra are 
    offset by a value of 0.2 relative to each other.  No Class I 
    spectra show evidence of any CO absorption.}
\end{figure}


\begin{thebibliography}{}
    
\bibitem[Adams, Lada, \& Shu(1987)]{ALS87} Adams, F. C., Lada, C. J., 
\& Shu, F. H. 1987, \apj, 213, 788 (ALS)

\bibitem[Andr\'{e} \& Montmerle(1994)]{AM94} Andr\'{e}, P., \& Montmerle, 
T. 1994, \apj, 420, 837

\bibitem[Biscaya Holzbach et al.(2000)]{BHCRL00} Biscaya Holzbach, A. M., 
Calvet, N. Rieke, G. H., \& Luhman, K. L. 2000, preprint

\bibitem[Boogert(1999)]{B99} Boogert, A. C. A. 1999, Ph.D. thesis, Groningen

\bibitem[Calvet, Hartmann, \& Strom(1997)]{CHS97} Calvet, N. Hartmann, L. \& 
Strom, S. E. 1997, \apj, 481, 912

\bibitem[Casali \& Matthews(1992)]{CM92} Casali, M. M. \& Matthews 1992, 
\mnras, 258, 399

\bibitem[D'Antona \& Mazzitelli(1997)]{DM97} D'Antona, F., \& 
Mazzitelli, I. 1997, in ``Cool Stars in Clusters and Associations,'' 
eds. R. Pallavicini \& G. Micela, Mem. S. A. It., 68, n. 4

\bibitem[Edwards et al.(1993)]{ESHSHHAMPG93} Edwards, S., et al.  
1993, \aj, 106, 372

\bibitem[Greene et al.(1993)]{GTTC93} Greene, T. P., Tokunaga, A. T., 
Toomey, D. W., \& Carr,  J. C. 1993, \procspie, 1946, 313

\bibitem[Greene \& Lada(1996)]{PaperI} Greene, T. P., \& Lada, C. J. 1996, 
\aj, 112, 2184 (Paper I)

\bibitem[Greene \& Lada(1997)]{PaperII} Greene, T. P., \& Lada, C. J. 1997, 
\aj, 114, 2157 (Paper II)

\bibitem[Greene et al.(1994)]{GWAYL94} Greene, T. P., Wilking, B. A., 
Andr\'{e}, P., Young, E. T., \& Lada, C. J. 1994,\apj, 434, 614
    
\bibitem[Hanner, Brooke, \& Tokunaga(1995)]{HBT95} Hanner, M. S., Brooke, 
T. Y., \& Tokunaga, A. T. 1995, \apj, 438, 250

\bibitem[Hanner, Tokunaga, \& Geballe(1992)]{HTG92} Hanner, M. S., Tokunaga, 
A. T., \& Geballe, T. R. 1992, \apjl, 395, L111

\bibitem[Herbst et al.(2000)]{HRHC00} Herbst, W., Rhode, K. L., 
Hillenbrand, L. A., \& Curran, G. 2000, 119, 261

\bibitem[Kenyon et al.(1998)]{KBTB98} Kenyon, S. J., Brown, D. I., Tout, C. 
A., Berlind, P. 1998, \aj 115, 2491

\bibitem[Luhman \& Rieke(1999)]{LR99} Luhman, K. L., \& Rieke, G. H.  1999, 
\apj, 525, 440

\bibitem[Montmerle et al.(2000)]{MGTK00} Montmerle, T., Grosso, N., Tsuboi, 
Y., \& Koyama, K. 2000, \apj, in press

\bibitem[Sekimoto et al.(1997)]{STUKTH97} Sekimoto, Y., Tatematsu, K., 
Umemoto, T., Koyama, K., Tsuboi, Y., \& Hirano, N. 1997, \apjl, 489, L63

\bibitem[Shu(1991)]{S91} Shu, F. H. 1991, in {\it The Physics of Star 
Formation and Early Stellar Evolution}, eds.  C. J. Lada \& N. D. 
Kylafis (Kluwer, Dodrecht), 365

\bibitem[Stahler(1988)]{S88} Stahler, S. W. 1988, \apj, 332, 804

\bibitem[Stassun et al.(1999)]{SMMV99} Stassun, K. G., Mathieu, R. 
D., Mazeh, T., \& Vrba, F. 1999, \aj, 117, 2941

%\bibitem[Strom, Kepner, \& Strom 1995]{} Strom, K. M., Kepner, J., \& 
%Strom, S. E. 1995, \apj, 438, 813

\bibitem[Tokunaga et al.(1990)]{TTCHE90} Tokunaga, A. T., Toomey, D. W., 
Carr, J. S., Hall, D. N. B., \& Epps, H. W. 1990, \procspie, 1235, 131

\bibitem[Wilking, Lada, \& Young(1989)]{WLY89} Wilking, B. A., 
Lada, C. J., \& Young, E. T. 1989, \apj, 340, 823 (WLY)

\end{thebibliography}
\end{document}